\documentclass[10pt]{article}

\usepackage{amsmath}
\usepackage{amssymb}

\usepackage{graphicx}

\usepackage{subfigure}

\usepackage{cite}

\usepackage{color} 

\usepackage[labelfont=bf,labelsep=period,justification=raggedright]{caption}

\makeatletter
\renewcommand{\@biblabel}[1]{\quad#1.}
\makeatother

\date{}

\begin{document}

\begin{flushleft}
{\Large
\textbf{The Anatomy of the Facebook Social Graph}
}
\\
Johan Ugander$^{1,2\ast}$, 
Brian Karrer$^{1,3\ast}$, 
Lars Backstrom$^{1}$, 
Cameron Marlow$^{1\dagger}$
\\
{\bf1} Facebook, Palo Alto, CA, USA \\
{\bf2} Cornell University, Ithaca, NY, USA \\
{\bf3} University of Michigan, Ann Arbor, MI, USA \\
$\ast$ These authors contributed equally to this work. \\
$\dagger$ Corresponding author: cameron@fb.com
\end{flushleft}

\section*{Abstract}

We study the structure of the social graph of active Facebook users, 
the largest social network ever analyzed. We compute numerous features 
of the graph including the number of users and friendships, the degree distribution, 
path lengths, clustering, and mixing patterns.  Our results center around three main 
observations.  First, we characterize the global structure of the graph, determining that the
social network is nearly fully connected, with 99.91\% of individuals belonging 
to a single large connected component, and we confirm the `six degrees of separation'
phenomenon on a global scale.  Second, by studying the average local clustering coefficient and 
degeneracy of graph neighborhoods, we show that while the Facebook graph as a whole is 
clearly sparse, the graph neighborhoods of users contain surprisingly dense structure.
Third, we characterize the assortativity patterns present in the graph 
by studying the basic demographic and network properties of users.  We observe clear 
degree assortativity and characterize the extent to which `your friends have 
more friends than you'.  Furthermore, we observe a strong effect of age
on friendship preferences as well as a globally modular community structure 
driven by nationality, but we do not find any strong gender homophily.
We compare our results with those from smaller social networks and find 
mostly, but not entirely, agreement on common structural network characteristics.   

\section*{Introduction}

The emergence of online social networking services over the past decade has revolutionized how social scientists study the structure of human relationships~\cite{boyd07}. As individuals bring their social relations online, the focal point of the internet is evolving from being a network of documents to being a network of people, and previously invisible social structures are being captured at tremendous scale and with unprecedented detail. In this work, we characterize the structure of the world's largest online social network, Facebook, in an effort to advance the state of the art in the empirical study of social networks.

In its simplest form, a social network contains individuals as vertices and edges as relationships between vertices~\cite{WF94}. This abstract view of human relationships, while certainly limited, has been very useful for characterizing social relationships, with structural measures of this network abstraction finding active application to the study of everything from bargaining power~\cite{Bonacich87} to psychological health~\cite{bearman2004}. Moreover, social networks have been observed to display a broad range of unifying structural properties, including homophily, clustering, the small-world effect, heterogeneous distributions of friends, and community structure~\cite{Newman10Book,KleinbergBook}.

Quantitative analysis of these relationships requires individuals to explicitly detail their social networks.  Historically, studies of social networks were limited to hundreds of individuals as data on social relationships was collected through painstakingly difficult means.  Online social networks allow us to increase the scale and accuracy of such studies dramatically because new social network data, mostly from online sources, map out our social relationships at a nearly global scale.  Prior studies of online social networks include research on Twitter, Flickr, Yahoo! 360, Cyworld, Myspace, Orkut, and LiveJournal among others~\cite{Ahn07, mislove07, mislove08, kumar10, kwak10}.

The trend within this line of research is to measure larger and larger representations of social networks, including networks derived from email~\cite{kossinets2006}, telephony \cite{eagle2010}, and instant messaging~\cite{Leskovec08} traces. Two recent studies of Renren~\cite{jiang10} and MSN messenger~\cite{Leskovec08} included 42 million and 180 million individuals respectively. Network completeness is especially important in the study of online social networks because unlike traditional social science research, the members of online social networks are not controlled random samples, and instead should be considered biased samples.  While the demographics of these networks have begun to approach the demographics of the global population at large~\cite{Chang2010}, the most accurate representation of our social relationships will include as many people as possible. We are not there yet, but in this paper we characterize the entire social network of active members of Facebook in May 2011, a network then comprised of $721$ million active users. To our knowledge, this is the largest social network ever analyzed. 

Facebook has naturally attracted the attention of researchers in the past. Some of this research has been devoted towards understanding small subsets of the Facebook population, including in particular the social networks of university students~\cite{traud08, lewis08, traud11}. Other studies have analyzed communication patterns and activity amongst segments of the user population~\cite{golder07, viswanath09}. Then another thread of research has measured some large-scale network properties of the Facebook graph through sampling, crawling, and other methods to collect network data~\cite{Catanese11, gjoka10}. Notably, these methodologies have no way of distinguishing between active and stale accounts. Unlike these studies, we analyze the entire Facebook graph in anonymized form and focus on the set of active user accounts reliably corresponding to people.

We defined a user of Facebook as an active member of the social network if they logged into the site in the last $28$ days from our time of measurement in May 2011 and had at least one Facebook friend.  We note that this definition is not precisely Facebook's ordinary definition of active user, and therefore some of our statistics differ slightly from company statistics.  According to our definition of active, the population of active Facebook users consisted of around $n = 721$ million individuals at the time of our measurements.  For comparison, estimating that the world's population was around $6.9$ billion people in May 2011 means that the network includes roughly $10$ percent of the world's population.  Restricting the comparison to individuals age 13 or more and with access to the internet (the set of individuals eligible to have Facebook accounts) would put this percentage significantly higher.  There were $68.7$ billion friendship edges at the time of our measurements, so the average Facebook user in our study had around $190$ Facebook friends.  Our analysis predated the existence of `subscriptions' on Facebook, and in this work we study only reciprocal Facebook friendships.

We also analyzed the subgraph of $149$ million U.S. Facebook users.  Using population estimates from the U.S. Census Bereau for $2011$, there are roughly $260$ million individuals in the U.S. over the age of $13$ and therefore eligible to create a Facebook account. Within the U.S., the Facebook social network therefore includes more than half the eligible population.  This subpopulation had $15.9$ billion edges, so the average U.S. user was friends with around $214$ other U.S. users.  This higher average value may reflect Facebook's deeper adoption in the U.S. as of May 2011.

Our goals with characterizing Facebook's social network are two-fold. First, we aim to advance the collective knowledge of social networks and satisfy widespread curiosity about social relationships as embodied in Facebook.  Second, we hope to focus the development of graph algorithms and network analysis tools towards a realistic representation of these relationships.  Towards these goals, we provide an accurate description of Facebook's social network here.

\section*{Results}

In this section, we apply a wide variety of graph measures to the Facebook social network.  As the network is truly enormous, we utilize extensive computational resources to perform these measurements.  However, the focus of this paper is on the results of these measurements, and therefore we relegate discussions of our techniques to the Methods.

\subsection*{The Facebook Graph}

\begin{figure}[!t]
\begin{center}
\subfigure[]{%
\includegraphics{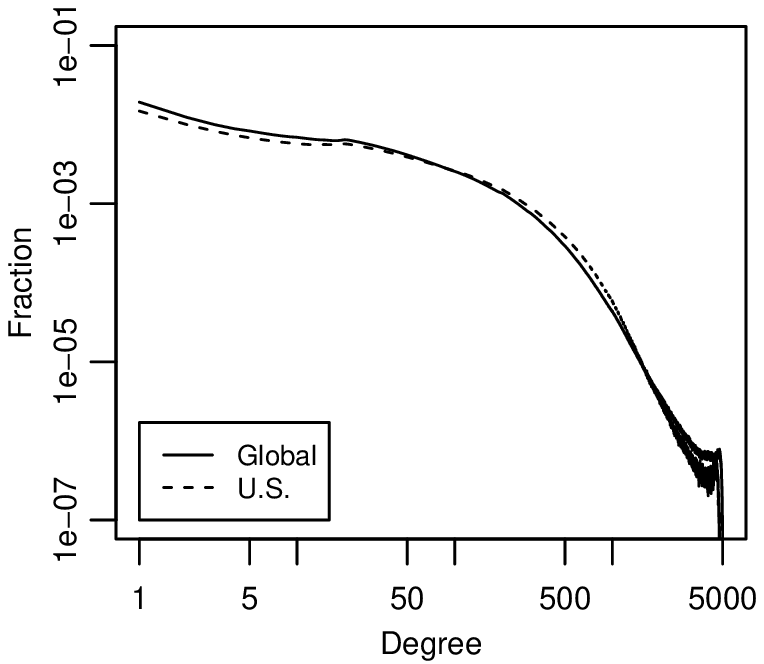}}
\subfigure[]{%
\includegraphics{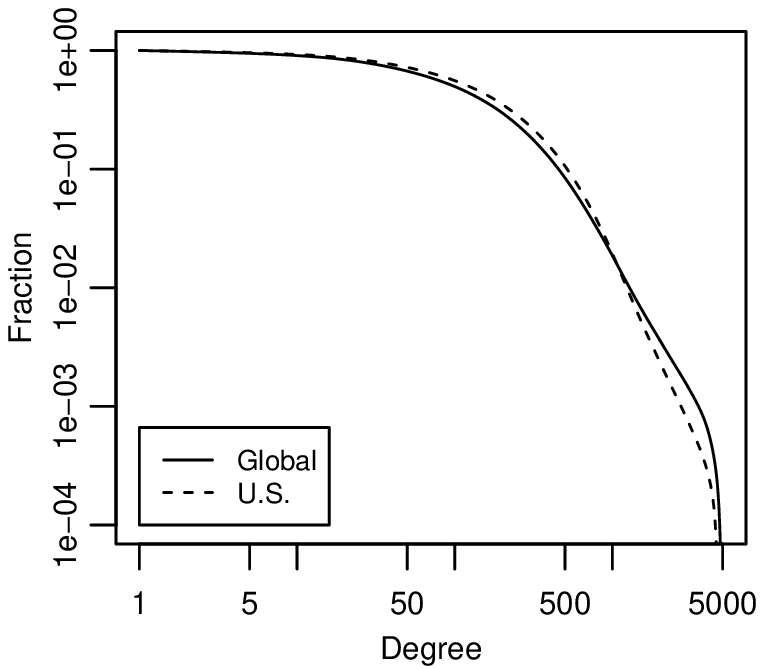}}
\end{center}
\caption{{\bf Degree distribution $p_k$.}  (a) The fraction of 
users with degree $k$ for both the global and U.S. population of
Facebook users.  (b) The complementary cumulative distribution function (CCDF).
The CCDF at degree $k$ measures the fraction of users who have degree $k$ or greater
and in terms of the degree distribution is $\sum_{k' \geq k} p_{k'}$.  For the U.S., 
the degree measures the number of friends also from the United States.}
\label{f:dd}
\end{figure}

{\bf Degree distribution}.
A fundamental quantity measured repeatedly in empirical studies of networks has 
been the \emph{degree distribution} $p_k$. The degree $k$ of an individual is 
the number of friends that individual has, and $p_k$ is the fraction of 
individuals in the network who have exactly $k$ friends. We computed the 
degree distribution of active Facebook users across the entire global population 
and also within the subpopulation of American users.  The global and U.S. degree 
distributions are shown in Fig.~\ref{f:dd}, displayed on a log-log scale.

Because the distribution for the U.S. is quite similar to that of the 
entire population, we focus our attention on the global degree distribution. 
The distribution is nearly monotonically decreasing, except for a small 
anomaly near $20$ friends. This kink is due to forces within the Facebook 
product to encourage low friend count individuals in particular to gain 
more friends until they reach $20$ friends. The distribution shows a clear 
cutoff at $5000$ friends, a limit imposed by Facebook on the number of 
friends at the time of our measurements.  Note that since $5000$ is 
nowhere near the number of Facebook users, each user is clearly friends with a 
vanishing fraction of the Facebook population. Reflecting most observed social 
networks, our social relationships are sparse.

Indeed, most individuals have a moderate number of friends on 
Facebook, less than $200$, while a much smaller population have 
many hundreds or even thousands of friends. The {\it median} friend count 
for global users in our study was $99$. The small population of users 
with abnormally high degrees, sometimes called \textit{hubs} in the 
networks literature, have degrees far larger than the average or median 
Facebook user. The distribution is clearly right-skewed with a high 
variance, but it is notable that there is substantial curvature 
exhibited in the distribution on a log-log scale. This curvature is 
somewhat surprising, because empirical measurements of networks have 
often claimed degree distributions to follow so-called power-laws, 
represented mathematically by $p_k \propto k^{-\alpha}$ for some 
$\alpha > 0$~\cite{BA99b, CSN09}.  Power-laws are straight lines on a log-log plot, 
and clearly the observed distribution is not straight.  We conclude, like
Ref.~\cite{gjoka10}, that strict power-law models are inappropriate 
for Facebook's degree distribution.  It is not our intent,
though, to determine which parametric form best models the 
distribution.  The relevant results are the monotonicity and curvature of
the degree distribution, the degrees of typical users, the large variance 
in degrees, and the network's sparsity. 

The sparsity of the network does not, however, imply that users are far 
from each other in Facebook's network.  While most pairs of users are not 
directly connected to each other, practically all pairs of users are 
connected via paths of longer lengths.  In the next section, we 
measure the distances between users in the social graph.

{\bf Path lengths}.
When studying a network's structure, the distribution of distances between 
vertices is a truly macroscopic property of fundamental interest. Here we 
characterize the {\it neighborhood functions} and the {\it average pairwise distances} 
of the Facebook and U.S. networks.

Formally, the neighborhood function $N(h)$ of a graph describes the number
of pairs of vertices $(u,v)$ such that $u$ is reachable from $v$ along a path
in the network with $h$ edges or less. Given the neighborhood function, 
the diameter of a graph is simply the maximum distance between
 any pair of vertices in the graph. The diameter is an extremal measure, 
 and it is commonly considered less 
interesting than the full neighborhood function, which measures what percentile 
of vertex pairs are within a given distance. The exact diameter can be wildly distorted 
by the presence of a single ill-connected path in some peripheral region of the 
graph, while the neighborhood function and its average are thought to robustly 
capture the `typical' distances between pairs of vertices.

Like many other graphs, the Facebook graph does not have paths between all pairs 
of vertices. This does not prevent us from describing the network using the 
neighborhood function though.  As we shall see in the next section, 
the vast majority of the network consists of one large connected component 
and therefore the neighborhood function is representative of the overwhelming 
majority of pairs of vertices.

\begin{figure}[!t]
\begin{center}
\includegraphics{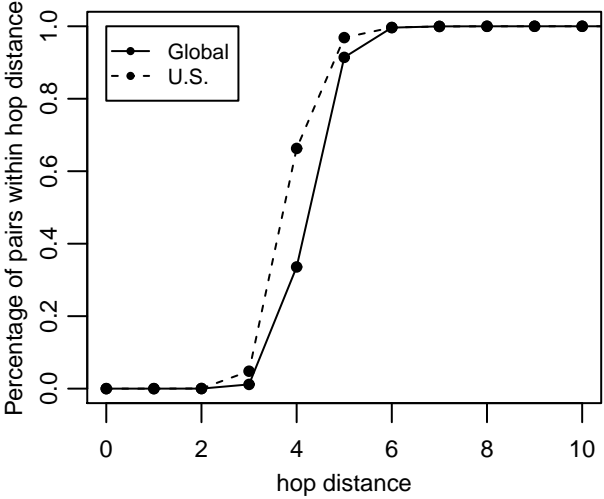}
\end{center}
\caption{{\bf Diameter.}  The neighborhood function $N(h)$ showing the percentage of 
user pairs that are within $h$ hops of each other. The average distance between users 
on Facebook in May 2011 was 4.7, while the average distance within the 
U.S. at the same time was 4.3.}
\label{f:diameter}
\end{figure}

Figure \ref{f:diameter} shows the neighborhood function computed for both the 
graph of all Facebook users as well as the graph of U.S. Facebook users, as of 
May 2011, using the recently developed HyperANF algorithm \cite{boldi2011}.
We find that the average distance between pairs of users was 4.7 for Facebook users 
and 4.3 for U.S. users. Short path lengths between individuals, the 
so-called ``six degrees of separation" found by Stanley Milgram's experiments 
investigating the social network of the United States \cite{Travers69}, are 
here seen in Facebook on a global scale. As Figure \ref{f:diameter} shows, fully 92\% 
of all pairs of Facebook users were within five degrees of separation, and 99.6\% 
were within six degrees. Considering the social network of only U.S. users, 96\% were 
within five degrees and 99.7\% were within six degrees. For the technical details 
behind these calculations, we refer the interested reader to a separate paper 
concerning the compression and traversal of the Facebook graph\cite{BackstromWWW}.

In order to show that these path length results are representative of the entire 
Facebook network, we now investigate the component structure of the graph.

{\bf Component sizes}.
Our conclusion from the previous section that the network has very
short average path lengths relies on the existence of such paths between
most pairs of vertices, a fact which we shall confirm in this section. We do so by finding the 
\textit{connected components} of the social network, where a connected 
component is a set of individuals for which each pair of individuals are connected by at least one 
path through the network.  Our neighborhood function calculations only 
computed distances between pairs of users within connected components because 
these are the only users actually connected via paths.  In order for the
results from the previous section to be interpreted as describing the diameter, we require that most, 
if not all, of the network be in one large connected component.

\begin{figure}[!t]
\begin{center}
\includegraphics{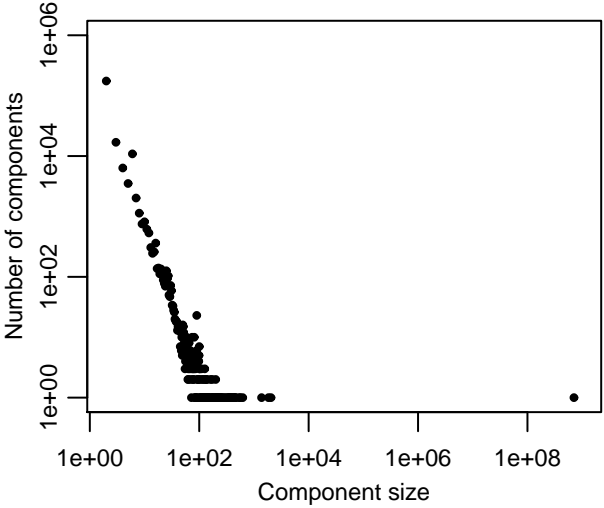}
\end{center}
\caption{{\bf Component size distribution.}  The fraction of 
components with a given component size on a log-log scale.  Most vertices
($99.91\%$) are in the largest component.}
\label{f:compsize}
\end{figure}

In Fig.~\ref{f:compsize}, we show the distribution of component sizes on 
log-log scales, found exactly
using an algorithm described in the Methods. While there are many connected components, 
most of these components are extremely small.  The second-largest connected component 
only has just over $2000$ individuals, whereas the largest connected component, the 
outlier all the way on the right-hand side of the figure, consists of $99.91\%$ percent of 
the network.  This component comprises the {\it vast} majority of active Facebook
users with at least one friend. So not only are the average path lengths between 
individuals short, these social connections exist between nearly everyone on 
Facebook.

The path lengths and component structure of the network give us a view
of the network at a macroscopic scale, and we now continue our investigation
by examining more local properties of the network.

{\bf Clustering coefficient and degeneracy}.
Earlier we characterized the number of friends per user by computing the degree
distribution, and we now perform a closer analysis of the social graph 
neighborhoods of users. The {\it neighborhood graph} for user $i$, sometimes 
called the {\it ego graph} or the {\it 1-ball}, is the vertex-induced subgraph 
consisting of the users who are friends with user $i$ and the friendships 
between these users. User $i$ is not included in their own neighborhood.

\begin{figure}[!tb]
\begin{center}
\subfigure[]{%
\includegraphics{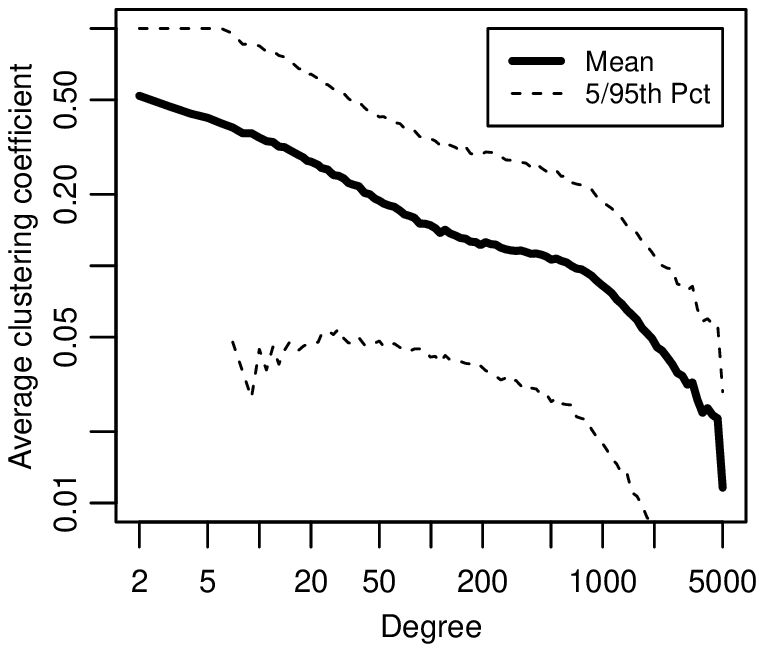}}
\subfigure[]{%
\includegraphics{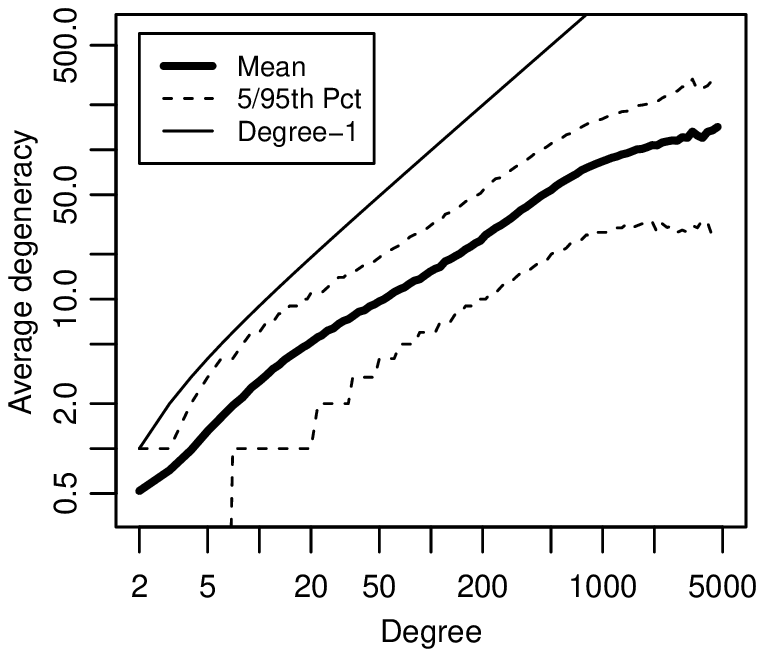}}
\end{center}
\caption{{\bf Local clustering coefficient and degeneracy.} The
clustering coefficient (a) and the degeneracy (b) as a function of 
degree on log-log scales. The means are shown as solid lines and the
dashed lines display the 5/95th percentiles. An upper-bound on the 
degeneracy given by degree minus 1 is also shown as the thin solid
line for comparison.}
\label{f:cc}
\end{figure}

We first computed the average {\it local clustering coefficient} for users as a 
function of degree, which for a vertex of degree $k$ measures the percentage 
of possible friendships between their $k$ friends (at most $k(k-1)/2$) are 
present in their neighborhood graph. This result is shown in Figure \ref{f:cc}a, where 
we note that the axes are log-log. 

We see that the local clustering coefficient is very large regardless of the
degree, compared to the percentage of possible friendships 
in the network as a whole, and more importantly, compared to measurements of
other online social networks.  For example, 
for users with 100 friends, the average local clustering coefficient is 0.14, 
indicating that for a median user, 14\% of all their friend pairs 
are themselves friends.  This is approximately {\it five times greater} 
than the clustering coefficient found in a 2008 study analyzing the 
graph of MSN messenger correspondences, for the same neighborhood 
size \cite{Leskovec08}. 

Meanwhile, our analysis also shows that the clustering coefficient decreases 
monotonically with degree, consistent with the earlier MSN messenger 
study and other studies. In particular, the clustering 
coefficient drops rapidly for users with close to 5000 friends, indicating that 
these users are likely using Facebook for less coherently social purposes 
and friending users more indiscriminately. 

Having observed such large clustering coefficients in local neighborhoods, we chose 
to study the sparsity of the neighborhood graphs further by measuring
their {\it degeneracy}. Formally, the degeneracy of an undirected graph 
$G$ is the largest $k$ for which $G$ has a non-empty $k$-core \cite{bollobas2001}.
Meanwhile, the {\it $k$-core} of a graph $G$ is the maximal subgraph of $G$
in which all vertices have degree at least $k$, or equivalently, the subgraph 
of $G$ formed by iteratively removing all vertices of degree less than $k$ 
until convergence.

The maximal $k$-core of a graph $G$ bears conceptual resemblance 
to the maximal $k$-clique of $G$, but it is important to note that a 
$k$-core is not necessarily a $k+1$-clique, unless the $k$-core contains 
exactly $k+1$ vertices. The $k$-core however offers a readily computable 
and robust indication of how tightly-knit a community exists within a given 
graph.

We report the average degeneracy as a function of user degree in Figure~\ref{f:cc}b, 
again plotted on a log-log scale.  Within the neighborhood graphs of users, we 
find that the average degeneracy is an increasing function of user degree.
This should be considered consistent with our expectations: the more friends you 
have, the larger a tight-knit community you are typically embedded within. What is 
however surprising is how dense these neighborhoods in fact are: for a user 
with 100 friends, the average degeneracy of their neighborhood is 15. 
Furthermore, for users with 500 friends, their average degeneracy is 53, 
meaning that they have at least 54 friends who all know 53 of their other 
friends. In contrast, Eppstein and Strash recently examined the degeneracy 
of several graphs, both social and non-social, and found that across the 
entire graphs (not examining only neighborhoods), the degeneracies were 
much more modest \cite{eppstein2011}. For the 36,692 vertex graph of Enron 
email communication, the degeneracy was only 43.  For the 16,706 vertex graph 
of arXiv astro-ph collaborations, the degeneracy was only 56. In contrast, we
find comparable degeneracies simply by considering the neighborhood of 
an average user with 500 friends. 

This suggests that even though the Facebook graph is sparse as a whole, 
when users accumulate sizable friend counts their friendships are far from 
indiscriminate, and instead center around sizable dense cores. We now consider 
the neighborhood of a vertex out to greater distances by examining the 
friends-of-friends of individual users.

\begin{figure}[!t]
\centering
\includegraphics{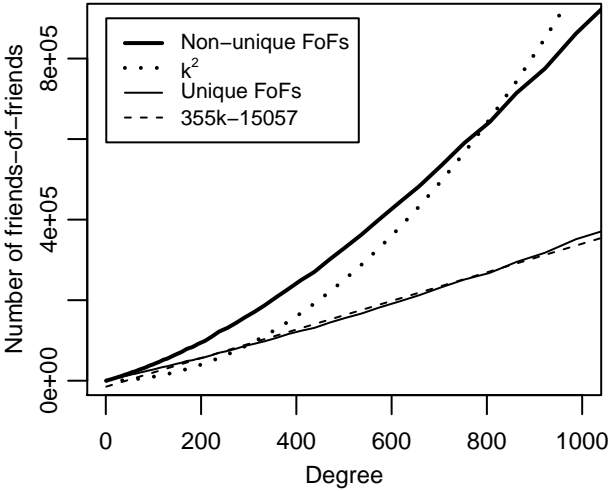} 
\caption{{\bf Friends-of-friends.}  The average number of unique and
non-unique friends-of-friends as described in the text as a function of
degree.  Degree squared and a linear fit of the unique friends-of-friends
 are shown for comparison as the dotted and dashed lines.}
\label{f:fofs}
\end{figure}

{\bf Friends of Friends}.  An important property of graphs to consider when designing
algorithms is the number of vertices that are within two hops of an initial vertex. This property
determines the extent to which graph traversal algorithms, such as breadth-first search,
are feasible. In Figure \ref{f:fofs}, we computed the average count of both unique and 
non-unique friends-of-friends as a function of degree. The non-unique friends-of-friends count 
corresponds to the number of length-two paths starting at an initial vertex and not
returning to that vertex.  The unique friends-of-friends count corresponds to the number of unique 
vertices reachable at the end of a length-two path.

A naive approach to counting friends-of-friends would assume that a user with $k$ friends has 
roughly $k^2$ non-unique friends-of-friends, assuming that their friends have 
roughly the same friend count as them. This could also be considered a generous 
naive estimate of the number of unique friends-of-friends, generous because we saw above that 
a significant fraction of your friends's friends are your friends. In reality, the number of 
non-unique friends of friends grows only moderately faster than linear, and the number of 
unique friends-of-friends grows very close to linear, with a linear fit producing a slope 
of 355 unique friends-of-friends per additional friend.

While the growth rate may be slower than expected, until a user has more than 800 friends, 
it's important to observe from the figure that the absolute amounts are unexpectedly large:
a user with with 100 friends has $27,500$ unique friends-of-friends and $40,300$ non-unique 
friends-of-friends. This is significantly more than the $100*99=9,900$ non-unique 
friends-of-friends we would have expected if our friends had roughly the same number 
of friends as us. This excess is related to a principle which we will discuss at length 
below, where we show the extent to which `your friends have more friends than you', an 
established result from prior studies of social networks~\cite{Feld91}.

\begin{figure}[!t]
\begin{center}
\subfigure[]{%
\includegraphics{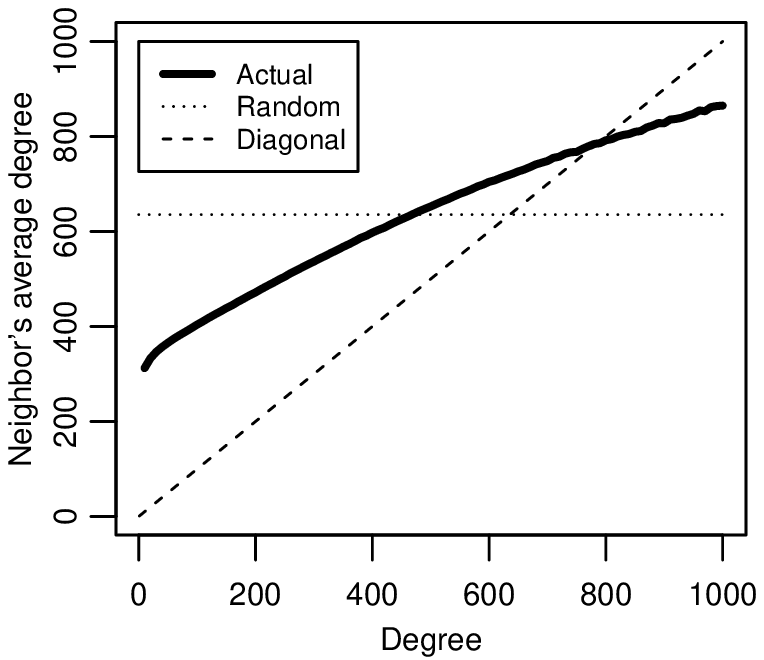}}
\subfigure[]{%
\includegraphics{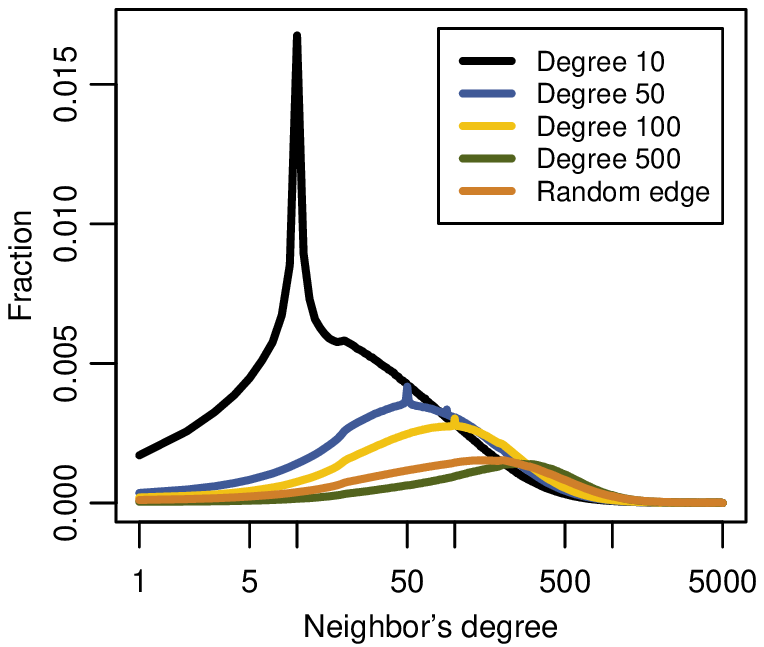}}
\end{center}
\caption{{\bf Degree correlations.} (a) The average neighbor degree of 
an individual with degree $k$ is the solid line.  The horizontal dashed 
line shows the expected value if there were no degree correlations in 
the network $\left\langle k^2 \right\rangle/\left\langle k \right\rangle$,
and the diagonal is shown as a dashed line. (b) The conditional probability 
$p(k'|k)$ that a randomly chosen neighbor of an individual with degree 
$k$ has degree $k'$. The solid lines, on the linear-log scale, show the 
measured values for four distinct degrees $k$ shown in the caption.  The 
orange line shows the expected distribution, $\frac{k' p_{k'}}{\left\langle 
k \right\rangle}$, if the degrees were uncorrelated.}
\label{f:neighbordegcombo}
\end{figure}

{\bf Degree correlations}.
The number of friendships in your local network neighborhood depends
on the number of friends, the degree, of your friends. In many  
social networks, online and offline, it has been noticed that your 
neighbor's degree is correlated with your own degree: it tends to be large
when your degree is large, and small when your degree is small, so-called 
\textit{degree assortativity}.  We can quantify these degree correlations 
by computing the Pearson correlation coefficient $r$ between degrees at 
the end of an edge~\cite{Newman02f, Newman03c}.  For the Facebook network, 
$r = 0.226$, displaying positive correlations with similar magnitude to other social graphs. 
This value is consistent with earlier studies of smaller networks including academic coauthorship and 
film actor collaborations, where values of $r$ range from $0.120$ to $0.363$ \cite{Newman02f}. Another more 
detailed measure $\left\langle k_{nn}\right\rangle (k)$, the average number 
of friends for a neighbor of an individual with $k$ friends~\cite{PVV01}, is shown
as the solid line in Fig.~\ref{f:neighbordegcombo}a.  (We use the notation
$\left\langle x \right\rangle$ to represent an average of a quantity $x$).  
The expected number of friends at the end of a randomly chosen edge, 
$\left\langle k^2 \right\rangle/\left\langle k \right\rangle = 635$, 
is shown as a horizontal dotted line.  Unlike this constant 
value --- which is our expectation if there were no degree correlations --- the solid 
line increases from near $300$ for low degree individuals to nearly 
$820$ for individuals with a thousand friends confirming the 
network's positive assortativity. (The measurements become noisy 
past $1000$ and we cut the figure off at this point for clarity.)

Comparing the solid line to the diagonal dashed line shows that until
you have nearly $700$ friends, your (average) neighbor has more
friends than you.  This phenomena has been discussed at length by
Feld~\cite{Feld91}, and Facebook displays the effect on a grand
scale. The fact that our average neighbors have so many more 
friends also explains why our naive friend-of-friend estimates in the previous 
section were far too low.

Feld's observation that `your friends have more friends than
you' is an important psychological paradox, applying 
to friendship as well as sexual partners. When people compare 
themselves to their friends, it is conceptually more appropriate to 
frame the comparison relative to the median of their friends, 
psychologizing the question as a matter of asking what one's 
`class rank' is amongst one's peers \cite{zuckerman2001}. Our 
finding with regard to the median is therefore perhaps more 
significant: we observe that 83.6\% of users have less 
friends than the median friend count of their friends. All these
individuals experience that more than half of their friends
have more friends than they do. 
For completeness, 
we also note that 92.7\% of users have less friends than the 
average friend count of their friends. 

However, we can do more than measure these simple statistics
and we characterize the conditional probability $p(k'|k)$ that 
a randomly chosen friend of an individual with degree $k$ has 
degree $k'$~\cite{MSZ04}.  We computed this for evenly spaced 
values of $k$, all multiples of ten, and show the distribution 
for a few example values of $k$ in Fig.~\ref{f:neighbordegcombo}b, 
along with the distribution if there were no degree correlations, 
i.e. the distribution of degrees found by following a randomly
selected edge.

First, note that the horizontal axis is log while the vertical
axis is linear.  Agreeing with Fig.~\ref{f:neighbordegcombo}a, the mean
of these distributions is clearly less than the mean of the orange distribution
which represents following a random edge in the network, except for the green 
line denoting $k=500$.  Again, the distributions shift
to the right as $k$ increases demonstrating the degree assortativity.
Furthermore, barring any strange non-smooth behavior between
the sampled values of $k$, the median for $p(k'|k)$ is greater than 
$k$ up until between $390$ and $400$ friends, confirming that the behavior of the
mean in Fig.~\ref{f:neighbordegcombo}a was not misleading.  Another observation 
from the figure, and data for other values of $k$ not shown, is that the 
modal degree of friends is exactly equal to $k$ until around $k = 120$.  So 
while your friends are likely to have more friends than you on average, the 
most likely number of your neighbor's friends is the same as your 
degree for low to moderate degree users.

\begin{figure}[!t]
\begin{center}
\subfigure[]{%
\includegraphics{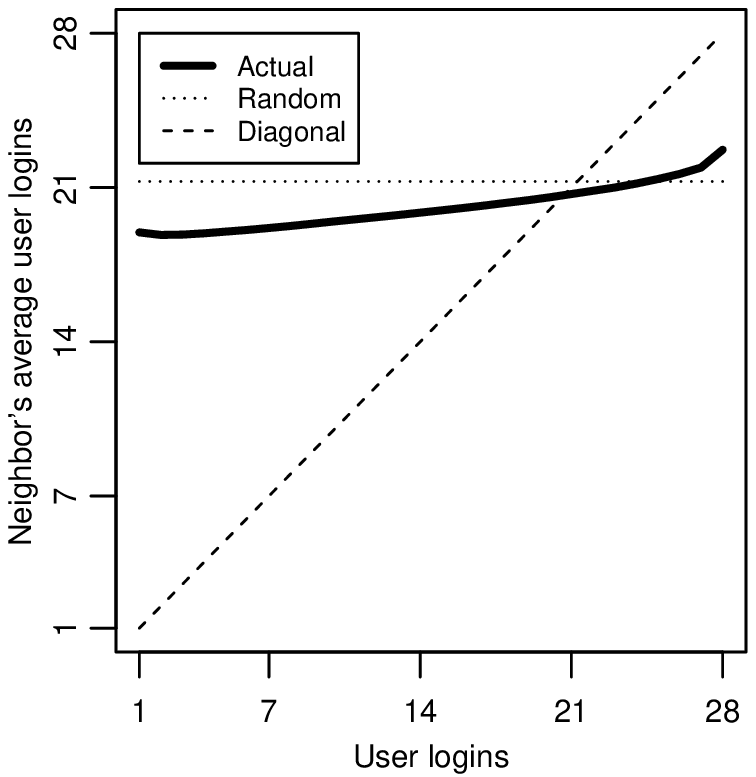}}
\subfigure[]{%
\includegraphics{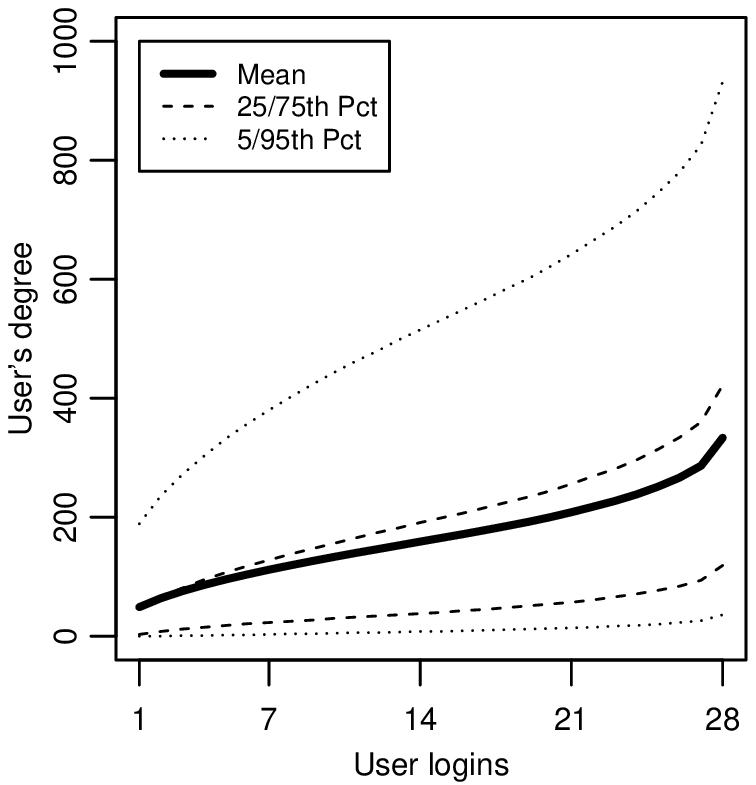}}
\end{center}
\caption{{\bf Login correlations.} (a) Neighbor's logins versus user's logins 
to Facebook over a period of $28$ days.  The solid line shows the actual mean
values and the horizontal line shows the average login value found by
following a randomly chosen edge.  The dashed line shows the diagonal. (b) A user's 
degree versus the number of days a user logged into Facebook in the $28$ day period. 
The solid line shows the mean user degree, the dashed lines the $25/75$ percentiles, 
and the dotted lines the $5/95$th percentiles.}
\label{f:logincombo}
\end{figure}

{\bf Site engagement correlation}.
Besides for degree correlations, we also examined correlations
amongst traits of individuals and network structure~\cite{MSC01}. We
now repeat our correlation calculations using the number of days users 
logged in during the 28-day window of the study, instead of degree, 
seen in Fig.~\ref{f:logincombo}a.  Again, we provide the 
average value at the end of a randomly selected edge and the 
diagonal line for comparison.

Unlike the degree case, here there
is an ambiguity in defining a random neighbor and hence the
average number of neighbor logins.  Our definition of 
random neighbor of vertices with trait $x$ is to first
select a vertex with trait $x$ in proportion to 
their degree and then select an edge connected to that 
vertex uniformly at random.  In other words, we give each 
edge connected to vertices with trait $x$ equal weight.  So
a vertex who is connected to $5$ vertices with trait $x$ is
given $5$ times as much weight in the average
as a vertex who connects to a single vertex with trait $x$.

Like the degree, your neighbor's site engagement is 
correlated with your site engagement, but the average number of neighbor 
logins is better represented by the horizontal random expectation 
than what was seen in the degree case.  The more interesting observation, though, is 
that the solid value is far larger than the diagonal value over most of the 
range from logging in $0$ to $20$ times in the past $28$ days.  So by the 
same line of reasoning for the degree case, up until you log in around 70 percent 
of days in a month, on average, your friends log into Facebook more than you do. 

We can understand this phenomena by examining
the correlation between an individual's degree and logging into 
Facebook.  A Facebook user provides and receives content 
through status updates, links, videos, and photos, etc. to and 
from their friends in the social network, and hence may be more 
motivated to log in if they have more friends.  Such a positive
correlation does exist between degree and logins, and we show that in
Fig.~\ref{f:logincombo}b.  A user who logs in more generally has more
friends on Facebook and vice versa.  So since your friends have
more friends than you do, they also login to Facebook more than
you do.

{\bf Other mixing patterns}.
There are many other user traits besides logging into Facebook
that can be compared to the network structure.  We focus on three other
such quantities with essentially complete coverage
for Facebook's users; age, gender, and country of origin, and 
characterize their homophily~\cite{MSC01} and mixing 
patterns~\cite{Newman03c}.

\begin{figure}[!t]
\centering
\includegraphics{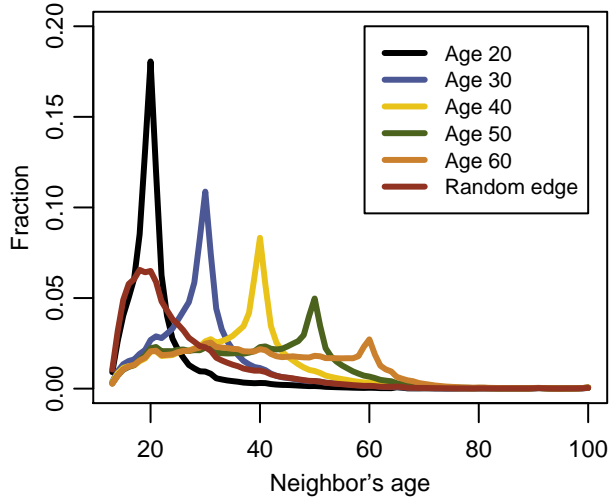} 
\caption{{\bf The distribution $p(t'|t)$ of ages $t'$ for the neighbors
of users with age $t$.}  
The solid lines show the measured distributions against the age $t$ 
described in the legend, and the red line shows the distribution 
of ages found by following a randomly chosen edge in the network.}
\label{f:agedist}
\end{figure}

We start by considering friendship patterns amongst individuals with
different ages, and compute the conditional probability $p(t'|t)$ of
selecting a random neighbor of individuals with age $t$ who has age
$t'$.  Again, random neighbor means that each edge connected to a vertex 
with age $t$ is given equal probability of being followed.  We display 
this function for a wide range of $t$ values in
Fig.~\ref{f:agedist}.  The resulting distributions are
not merely a function of the magnitude of the age difference $|t-t'|$ as might
naively be expected, and instead are asymmetric about a maximum value
of $t'=t$.  Unsurprisingly, a random neighbor is most likely to be the same age as
you. Less obviously, the probability of friendship with older individuals 
falls off rapidly, nearly exponentially, from the mode.  Below the mode, 
the distributions also fall off, but then level out to a value that is 
nearly independent of the user's age $t$ (see for example the blue, 
yellow and green lines).  And from the figure we notice that as $t$ increases 
the variance in the distribution increases.  Roughly speaking, younger individuals 
have most of their friends within a small age range while older individuals
have a much wider range.  None of this behavior is evident when comparing 
to the distribution of ages at the end of a randomly chosen edge, the red line, 
which is centered around $20$. So while it is obvious that age matters to our 
social relationships, the Facebook social network shows non-trivial asymmetric 
patterns, consistent across user ages $t$.

Switching to gender, we compute the conditional probability $p(g'|g)$
that a random neighbor of individuals with gender $g$ has gender $g'$ 
where we denote male by $M$ and female by $F$.  For friends of male users, 
we find that $p(F|M) = 0.5131$ and $p(M|M)=0.4869$.
For friends of female users, we find that $p(F|F) = 0.5178$ and 
$p(M|F) = 0.4822$.  In both cases, we see that a random neighbor
is more likely to be female.  

In order to understand this result, we compare to the probability of 
following a randomly selected edge and arriving at a particular gender.  These 
probabilities are given by $p(F) = 0.5156$ and $p(M) = 0.4844$ respectively.  The 
probability is higher for females because the number of edge ends, called 
stubs in the networks literature, connected to females is higher than for 
males.  While there are roughly $30$ million fewer active female users 
on Facebook, the average female degree ($198$) is larger than the average 
male degree ($172$), resulting in $p(F) > p(M)$.

Comparing these quantities, we see that $p(F|M) < p(F) < p(F|F)$ and 
$p(M|F) < p(M) < p(M|M)$.  However, the magnitude of the difference between
these probabilities is extremely small and only differs in the 
thousandths place. So if there is a preference for same gender 
friendships on Facebook, the effect appears minimal at most. 

Lastly, we turn to country of origin, a categorical variable divided into
$249$ categories according to the ISO 3166-1 country code standard. 
These labels are attributed to users based on the user's most recent 
IP address login source and known correspondences between IP addresses
and geographic locations. While imperfect, so-called geo-IP data is generally
reliable on a national level.

Intuitively, we expect to have many more friends from
our country of origin then from outside that country, and the data
shows that $84.2\%$ percent of edges are within countries.  So the network
divides fairly cleanly along country lines into network \textit{clusters} or
\textit{communities}. This mesoscopic-scale organization is to be expected
as Facebook captures social relationships divided by national borders.  
We can further quantify this division using the modularity
$Q$~\cite{NG04} which is the fraction of edges within communities minus
the expected fraction of edges within communities in a randomized version 
of the network that preserves the degrees for each individual~\cite{MR95}, 
but is otherwise random.  In this case, the communities are
the countries. The computed value is $Q=0.7486$ which is quite large~\cite{Fortunato10} 
and indicates a strongly modular network structure at the scale of countries.
Especially considering that unlike numerous studies using the modularity to detect 
communities, we in no way attempted to maximize it directly, and instead merely 
utilized the given countries as community labels.

\begin{figure}[!t]
\begin{center}
\includegraphics[width=0.68 \textwidth]{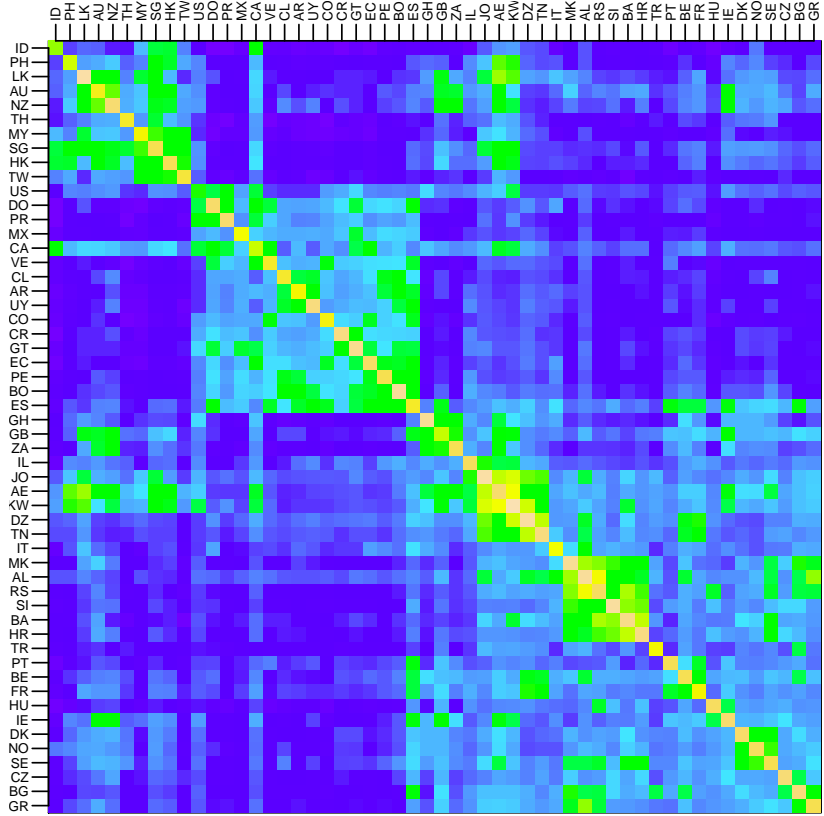}
\end{center}
\caption{{\bf Normalized country adjacency matrix.}
Matrix of edges between countries with $>1$ million users and $>50$\% Facebook 
penetration shown on a log scale. To normalize, we divided each element of the
adjacency matrix by the product of the row country degree and column country
degree. 
}
\label{f:country}
\end{figure}

We visualize this highly modular structure in Fig.~\ref{f:country}.
The figure displays a heatmap of the number of edges between the $54$ countries where 
the active Facebook user population exceeds one million users and is more 
than 50\% of the internet-enabled population~\cite{internetstats}.  To be entirely
accurate, the shown matrix contains each edge twice, once in both directions, and
therefore has twice the number of edges in diagonal elements.  The number of 
edges was normalized by dividing the $ij$th entry by the row and column sums, 
equal to the product of the degrees of country $i$ and $j$.  The ordering 
of the countries was then determined via complete linkage hierarchical clustering.  

While most of the edges in the figure are on the diagonal, the log-scale clearly 
highlights further modular structure in friendships between countries. The countries 
fall into groups, the clearly square-like patterns in the matrix, with 
preferential friendship patterns amongst citizens of different countries.  

The complete list of countries in the order presented in the matrix is shown in
Table~\ref{tab:country}.  Many of the resulting country groupings are
intuitive according to geography.  For example, there are clear groups corresponding
to the South Pacific, North and Central America, South America, North Africa
and the Middle East, Eastern Europe and the Mediterranean, and to
the Nordic countries of Denmark, Sweden, and Norway.  Other more curious
groupings, not clearly based on geography, include the combination of 
the United Kingdom, Ghana, and South Africa, which may reflect strong historical ties.
The figure clearly demonstrates that not only are friendships predominantly 
between users within the same country, but that friendships between countries 
are also highly modular, and apparently influenced by geography.  The influence 
of geographical distance on friendship has been previously discussed \cite{LibenNowell2005,Barthelemy2010}, 
but earlier work on Facebook has not examined the country-scale structure of friendships \cite{Backstrom2010}.  
Speculating, some groupings of countries may be better explained by historical and cultural relations than
simple geographical distance.

\section*{Discussion}

In this paper, we have characterized the structure of the Facebook 
social graph using many metrics and tools.  To our knowledge, 
our study is the largest structural analysis of a social network 
performed to date. 

We began by characterizing the degree distribution which was shown 
to be skewed with a large variance in friendship count.  Unlike 
many other networks, a pure power-law was seen to be 
inappropriate for the degree distribution of Facebook, although
hubs certainly exist.

The small-world effect and six degrees of separation were then
confirmed on a truly global scale.  The average distance between 
vertices of the giant component was found to be 4.7, and we interpret 
this result as indicating that individuals on Facebook have potentially
tremendous reach. Shared content only needs to advance a few 
steps across Facebook's social network to reach a substantial 
fraction of the world's population.

We have found that the Facebook social network
is nearly fully connected, has short average path lengths, and high clustering.
Many other empirical networks, social and non-social, have also been observed with
these characteristics, and Watts and Strogatz called networks with these 
properties `small-world networks'~\cite{WS98}.

Because our friends have more friends than we do,
individuals on Facebook have a surprisingly large number of
friends-of-friends. Further, our friends are highly clustered
and our friendships possess dense cores, a phenomena not
noticed in smaller social networks. This neighborhood structure
has substantial algorithmic implications for graph traversal
computations.  Breadth-first search out to distance two 
will potentially query a large number of individuals 
due to positive degree correlation, and then will
follow many edges to individuals already found 
due to the clustering.

We also performed an exploratory comparison
of the network structure with user traits including
login behavior, age, gender, and country.  We found
interesting mixing patterns, including that 
`your friends login to Facebook more than you do'
and a strong preference for friends of around the
same age and from the same country.

Community structure was shown to be clearly evident in
the global network, at the scale of friendships between and within
countries. Countries in turn were seen to themselves exhibit a 
modular organization, largely dictated by geographical
distance. Unlike prior studies of networks, this community 
structure was discovered without much effort, indicating the 
significant structural insights that can be derived from
basic demographic information.

While our computations have elucidated many aspects of the structure of the 
world's largest social network, we have certainly not
exhausted the possibilities of network analysis.
We hope that this information is useful both for social science
research and for the design of the next-generation of graph
algorithms and social network analysis techniques.

\section*{Materials and Methods}

Unless otherwise noted, calculations were performed on a Hadoop cluster with 2,250 machines, using the Hadoop/Hive data analysis framework developed at Facebook \cite{thusoo2010,thusoo2010b}.

For analyzing network neighborhoods, 5,000 users were randomly selected using reservoir sampling for each of 100 log-spaced neighborhood sizes, creating a sample of 500,000 users for which the analysis was performed. The percentiles shown for the clustering coefficient and degeneracy for each neighborhood size are therefore empirical percentiles from this sample population of 5,000 users.

To analyze the component structure of the network, we used the Newman-Zipf (NZ)
algorithm~\cite{NZ01}.  The NZ algorithm, a type of Union-Find algorithm with
path compression, records component structure dynamically as edges are added
to a network that begins completely empty of edges.  When all edges are added, 
the algorithm has computed the component structure of the network.  Crucially for our 
purposes, the NZ algorithm does not require the edges to be retained in memory. We apply it to the 
Facebook network on a single computer with 64GB of RAM by streaming over the list of edges.

For path length calculations,  neighborhood functions were computed on a single 24-core machine with 72 GB of RAM using the HyperANF algorithm~\cite{boldi2011}, averaging across 10 runs. For the technical details behind this formidable computation, see \cite{BackstromWWW}.



\begin{table}
\parbox{.45\linewidth}{
\centering
\begin{tabular}{c|c}
\hline
Country & ISO code \\
\hline
Indonesia & ID \\
Philipines & PH \\
Sri Lanka & LK \\
Australia & AU \\
New Zealand & NZ \\
Thailand & TH \\
Malaysia & MY \\
Singapore & SG \\
Hong Kong & HK \\
Taiwan & TW \\
United States & US \\
Dominican Republic & DO \\
Puerto Rico & PR \\
Mexico & MX \\
Canda & CA \\
Venezuela & VE \\
Chile & CL \\
Argentina & AR \\
Uruguay & UY \\
Colombia & CO \\
Costa Rica & CR \\
Guatemala & GT \\
Ecuador & EC \\
Peru & PE \\
Bolivia & BO \\
Spain & ES \\
Ghana & GH \\
\hline
\end{tabular}
}
\hfill
\parbox{.45\linewidth}{
\centering
\begin{tabular}{c|c}
\hline
Country & ISO code \\
\hline
United Kingdom & GB \\
South Africa & ZA \\
Israel & IL \\
Jordan & JO \\
United Arab Emirates & AE \\
Kuwait & KW \\
Algeria & DZ \\
Tunisia & TN \\
Italy & IT \\
Macedonia & MK \\
Albania & AL \\
Serbia & RS \\
Slovenia & SI \\
Bosnia and Herzegovina & BA \\
Croatia & HR \\
Turkey & TR \\
Portugal & PT \\
Belgium & BE \\
France & FR \\
Hungary & HU \\
Ireland & IE \\
Denmark & DK \\
Norway & NO \\
Sweden & SE \\
Czech Republic & CZ \\
Bulgaria & BG \\
Greece & GR \\
\hline
\end{tabular}
}
\caption{ISO country codes used as labels in Figure~\ref{f:country}.}
\label{tab:country}
\end{table}

\end{document}